\PassOptionsToPackage{unicode}{hyperref}
\PassOptionsToPackage{hyphens}{url}
\documentclass[
]{article}
\usepackage{amsmath,amssymb}
\usepackage{iftex}
\ifPDFTeX
  \usepackage[T1]{fontenc}
  \usepackage[utf8]{inputenc}

\DeclareUnicodeCharacter{207B}{\textsuperscript{-}}
\DeclareUnicodeCharacter{2080}{\textsubscript{0}}
\DeclareUnicodeCharacter{2081}{\textsubscript{1}}
\DeclareUnicodeCharacter{2082}{\textsubscript{2}}
\DeclareUnicodeCharacter{2083}{\textsubscript{3}}
\DeclareUnicodeCharacter{2084}{\textsubscript{4}}
\DeclareUnicodeCharacter{2085}{\textsubscript{5}}
\DeclareUnicodeCharacter{2086}{\textsubscript{6}}
\DeclareUnicodeCharacter{2087}{\textsubscript{7}}
\DeclareUnicodeCharacter{2088}{\textsubscript{8}}
\DeclareUnicodeCharacter{2089}{\textsubscript{9}}
\DeclareUnicodeCharacter{03B8}{$\theta$}
\DeclareUnicodeCharacter{03BB}{$\lambda$}
\DeclareUnicodeCharacter{03BC}{$\mu$}
\DeclareUnicodeCharacter{03BE}{$\xi$}
\DeclareUnicodeCharacter{03C0}{$\pi$}
\DeclareUnicodeCharacter{03C1}{$\rho$}
\DeclareUnicodeCharacter{03C3}{$\sigma$}
\DeclareUnicodeCharacter{2206}{$\Delta$}
\DeclareUnicodeCharacter{2212}{--}
\DeclareUnicodeCharacter{221D}{$\propto$}
\DeclareUnicodeCharacter{223C}{$\sim$}
\DeclareUnicodeCharacter{2248}{$\approx$}
\DeclareUnicodeCharacter{2261}{$\equiv$}
\DeclareUnicodeCharacter{2265}{$\geq$}

  \usepackage{textcomp} 
\else 
  \usepackage{unicode-math} 
  \defaultfontfeatures{Scale=MatchLowercase}
  \defaultfontfeatures[\rmfamily]{Ligatures=TeX,Scale=1}
\fi
\usepackage{lmodern}
\ifPDFTeX\else
\fi
\IfFileExists{upquote.sty}{\usepackage{upquote}}{}
\IfFileExists{microtype.sty}{
  \usepackage[]{microtype}
  \UseMicrotypeSet[protrusion]{basicmath} 
}{}
\makeatletter
\@ifundefined{KOMAClassName}{
  \IfFileExists{parskip.sty}{%
    \usepackage{parskip}
  }{
    \setlength{\parindent}{0pt}
    \setlength{\parskip}{6pt plus 2pt minus 1pt}}
}{
  \KOMAoptions{parskip=half}}
\makeatother
\usepackage{xcolor}
\usepackage{longtable,booktabs,array}
\usepackage{calc} 
\IfFileExists{footnotehyper.sty}{\usepackage{footnotehyper}}{\usepackage{footnote}}
\makesavenoteenv{longtable}
\usepackage{graphicx}
\usepackage{geometry}
\makeatletter
\def\maxwidth{\ifdim\Gin@nat@width>\linewidth\linewidth\else\Gin@nat@width\fi}
\def\maxheight{\ifdim\Gin@nat@height>\textheight\textheight\else\Gin@nat@height\fi}
\makeatother
\setkeys{Gin}{width=\maxwidth,height=\maxheight,keepaspectratio}
\makeatletter
\def\fps@figure{htbp}
\makeatother
\ifLuaTeX
  \usepackage{selnolig}  
\fi
\usepackage{bookmark}
\IfFileExists{xurl.sty}{\usepackage{xurl}}{} 
\hypersetup{
  hidelinks,
  pdfcreator={LaTeX via pandoc}}

\author{}
\date{}

\begin{document}

\textbf{Direct Observation of Chemical Short-Range Order in CoCrNi Alloy
Using Neutron Diffraction}

Vinícius P. Bacurau\textsuperscript{a,b}, Camilo
Salvador\textsuperscript{c}, Guilherme C. Stumpf\textsuperscript{a,b},
Angelo F. Andreolli\textsuperscript{a}, Caroline B.
Stoco\textsuperscript{a,b}, Eric M. Mazzer\textsuperscript{a,b}, Lewis
Owen\textsuperscript{d}, Yifan Cao\textsuperscript{e}, Rodrigo
Freitas\textsuperscript{e}, Daniel Miracle\textsuperscript{f}, Francisco
G. Coury\textsuperscript{a,b}

\emph{\textsuperscript{a}Department of Materials Engineering, Federal
University of São Carlos (UFSCar), São Carlos, Brazil}

\emph{\textsuperscript{b}Graduate Program in Materials Science and
Engineering, Federal University of São Carlos, São Carlos, São Paulo,
Brazil}

\emph{\textsuperscript{c}Université Paris-Saclay, CEA, Service de
recherche en Corrosion et Comportement des Matériaux, SRMP, 91191 Gif
Sur Yvette, France}

\emph{\textsuperscript{d}Department of Materials Science and
Engineering, University of Sheffield, Sheffield, United Kingdom}

\emph{\textsuperscript{e}Department of Materials Science and
Engineering, Massachusetts Institute of Technology, Cambridge, MA, USA}

\emph{\textsuperscript{f}AF Research Laboratory, Materials and
Manufacturing Directorate, Wright-Patterson AFB, OH, USA}

Corresponding authors at: \emph{Department of Materials Engineering,
Federal University of São Carlos (UFSCar), São Carlos, Brazil, Graduate
Program in Materials Science and Engineering, Federal University of São
Carlos, São Carlos, São Paulo, Brazil}

Corresponding authors

E-mail addresses:
\href{mailto:fgcoury@ufscar.br}{\nolinkurl{fgcoury@ufscar.br}} (F.G
Coury) and
\href{mailto:vinicius.bacurau@estudante.ufscar.br}{\nolinkurl{vinicius.bacurau@estudante.ufscar.br}}
(V.P. Bacurau)

Abstract:

This study provides experimental evidence of chemical short-range order
(CSRO) in the equiatomic CoCrNi alloy, identified through neutron
diffraction. The phenomenon manifests as a distinct diffuse peak at Q $\approx$
1.85 \AA$^{-1}$, the intensity increases under thermodynamically favorable
conditions for CSRO development such as prolonged aging (100 h and 240
h) at 748 K or shorter aging (24 h) at slightly higher temperature (798
K). The degree of ordering was measured by integrating the diffuse
scattering intensity, revealing that the gas-atomized sample, i.e. the
sample with the least amount of CSRO, still displays approximately 70\%
of the CSRO level observed in the sample subsequently aged for 240 h at
748 K, i.e. the sample with the highest amount of CSRO produced in this
study. Predictive atomistic simulations reproduced both the presence and
position of the diffuse peak, while two-dimensional fast Fourier
transform (FT-2D) analyses indicated that reflections at (1 ½ 0) within
the \textless001\textgreater zone axis originate from some structural
projections associated with like D0\textsubscript{22},
Pt\textsubscript{2}Mo and D1a motifs. Complementary small-angle neutron
scattering (SANS) measurements identified Ni-rich, disk-shaped domains
with radii of approximately 11 \AA and thicknesses of about 1 \AA,
consistent with nanoscale CSRO characteristic length scale. These
findings demonstrate that CSRO is an intrinsic and energetically
favorable feature of the CoCrNi system, remaining stable even under
rapid solidification and further enhanced by low-temperature aging.
Combined use of neutron diffraction and atomistic modeling provides a
framework for probing local ordering phenomena in multi-principal
element alloys (MPEAs).


\begin{enumerate}
\def\labelenumi{\arabic{enumi}.}
\item
  \textbf{Introduction}
\end{enumerate}

The development of multi-principal element alloys (MPEAs) has sparked
interest in understanding their local atomic configurations,
particularly the phenomenon of chemical short-range order (CSRO)
{[}1--5{]}. Among these alloys, the equiatomic CoCrNi solid solution has
emerged as a model system due to its exceptional mechanical properties
and microstructural stability at cryogenic and ambient temperatures
{[}6--8{]}.

CSRO refers to the non-random arrangement of atoms in a lattice within a
few nearest-neighbor shells in otherwise chemically disordered solid
solutions. While its presence is well established in conventional binary
alloys {[}9--12{]}, its detection in concentrated multicomponent systems
remains challenging, especially when the constituent elements are
adjacent in the periodic table and exhibit limited scattering or weak
contrast in conventional X-ray or electron-based techniques
{[}13--15{]}.

Several studies have reported the detection of CSRO through diffuse
peaks observed in transmission electron microscopy (TEM) diffraction
patterns {[}3,4,14{]}. However, Coury et al. {[}16{]} demonstrated that
many of the diffuse reflections commonly attributed to ordering
phenomena in selected area electron diffraction (SAED) patterns of
face-centered cubic (FCC) alloys may actually originate from
higher-order Laue zones. This finding highlights the importance of
caution when interpreting CSRO signatures based solely on electron
diffraction.

Along similar lines, Walsh \emph{et al.} {[}17{]} reported that features
historically identified as indicators of CSRO in FCC alloys may instead
originate from alternative mechanisms, including thermal and static
atomic displacements, lattice vibrations, and strain fields, as well as
secondary effects such as planar defects, surface contributions, and
dynamic scattering. These factors can generate diffuse spots that mimic
CSRO, reducing the reliability of electron diffraction as a primary tool
for detecting CSRO.

The selection of an appropriate diffraction technique to accurately
detect chemical order or disorder is strongly dependent on the alloy
composition, as the ideal radiation source is the one that provides the
highest contrast between the constituent elements.

In the case of equiatomic CrCoNi, widely regarded as a benchmark system
for CSRO studies, the atomic numbers of Cr, Co, and Ni are very similar
(24, 27, and 28 electrons, respectively). As a result, X-ray and
electron diffraction techniques lack sufficient contrast to reliably
distinguish between these elements, rendering them ineffective for
detecting CSRO in this alloy {[}15{]}. In contrast, neutron diffraction
offers a clear advantage. The coherent neutron scattering lengths of Cr,
Co, and Ni differ significantly (3.64 fm, 2.49 fm, and 10.3 fm,
respectively), providing enhanced chemical sensitivity. This increased
contrast enables the detection of subtle features associated with
chemical short-range order.

These limitations have driven the search for more robust approaches to
identify and measure CSRO in CrCoNi. In this context, neutron
diffraction has emerged as a particularly suitable technique, owing to
the weak dependence of neutron scattering lengths on atomic number and
the resulting improvement in chemical contrast among the constituent
elements{[}18{]}.

To overcome these challenges, Bacurau et al. {[}5{]} combined
high-precision differential scanning calorimetry and
machine-learning-assisted atomistic simulations to unambiguously detect
and quantify CSRO in CoCrNi and CrNi\textsubscript{2} alloys. Their
methodology allowed not only the identification of CSRO states with high
sensitivity but also the estimation of the enthalpy associated with CSRO
formation and dissolution. Notably, they showed that even in the absence
of long-range order, significant changes in calorimetric response could
be attributed to local ordering transitions.

CSRO in equiatomic CrCoNi has also been investigated through extended
X-ray absorption fine structure (EXAFS) and pair distribution function
(PDF) analyses {[}13,19,20{]}. Spin-polarized DFT (density functional
theory) simulations {[}21{]} have suggested that magnetic interactions
may drive the tendency to avoid Cr--Cr nearest neighbor pairs.
Furthermore, studies on binary Ni--Cr alloys with dilute Cr content have
revealed the presence of diffuse neutron scattering at half-integer
positions such as (1 ½ 0), consistent with the development of
short-range order involving Cr avoidance in first-neighbor shells.
Interestingly, however, this does not correspond to the typical
long-range ordered phases observed for Ni-Cr systems {[}22{]}.

Despite these improvements, the literature still lacks a direct and
consistent measure of CSRO levels in CrCoNi across distinct processing
conditions. To date, no study has unambiguously established the extent
of CSRO in this alloy system using diffraction techniques with
sufficient sensitivity to resolve differences between rapidly solidified
and thermally aged states. In the present work, we demonstrate the
presence of CSRO via diffuse diffraction spots at the (1 ½ 0) position
by combining neutron diffraction experiments with simulated diffraction
patterns generated from atomistic configurations obtained using accurate
machine-learning interatomic potentials {[}23{]}. These results provide
direct experimental evidence that even the rapidly solidified state
exhibits CSRO, albeit to a lesser degree than the thermally aged sample
{[}24{]}. This indicates that the emergence of CSRO is directly linked
to the attainment of lower free-energy configurations during the
solidification process.

\begin{enumerate}
\def\labelenumi{\arabic{enumi}.}
\setcounter{enumi}{1}
\item
  \textbf{Experimental procedures}
\end{enumerate}

\textbf{2.1 Alloy production}

An equiatomic CrCoNi alloy was synthesized by inert gas atomization
using argon (Ar), a process that ensures ultrafast solidification rates,
employing commercially pure elements (≥ 99.9\%). The resulting powders
were sealed under an inert atmosphere in high-purity quartz tubes,
annealed at 1373 K for 24 h, and subsequently water-quenched. Samples
were investigated in the as-atomized state, after homogenization, and
following various CSRO-aging treatments: 748 K for 100 h and 240 h, and
798 K for 24 h, also performed in encapsulated inert environments. All
heat treatments were followed by water quenching. The sealing of the
powders within quartz tubes prevented direct contact with water during
the quenching process, ensuring that the samples were not exposed to
oxidation or contamination. A summary of all experimental conditions
analyzed in this study is provided in Table 1. The selection of thermal
treatment temperatures and times were based on the work of Bacurau
\emph{et al.} {[}5{]}, in which the presence of CSRO was indirectly
detected.

Table 1: Nomenclature used throughout the text and thermal treatment
conditions for each sample.

\begin{center}
\begin{tabular}{p{0.32\textwidth}p{0.60\textwidth}}
\toprule
\textbf{Name used in the text} & \textbf{Condition of treatment} \\ 
\midrule
Gas atomized & As-atomized \\ 
Homogenized & Annealed at 1373 K for 24 h + water quenching \\ 
100 h/748 K & Homogenized and aged at 748 K for 100 h \\ 
240 h/748 K & Homogenized and aged at 748 K for 240 h \\ 
24 h/798 K & Homogenized and aged at 798 K for 24 h \\ 
\bottomrule
\end{tabular}
\end{center}

Microstructural characterization was performed using a scanning electron
microscope (SEM, FEG Tescan Mira3) equipped with energy-dispersive X-ray
spectroscopy (EDS) and electron backscatter diffraction (EBSD)
detectors, operating at an accelerating voltage of 25 kV. Sample
preparation for microstructural analysis involved sequential grinding
using SiC papers (220--1500 grit), followed by polishing with a 1 $\mu$m
alumina suspension. A final polishing step was performed for 12 h in a
colloidal silica solution using a Buehler VibroMet 2 polisher.

These analyses were carried out to evaluate the crystallographic texture
and compositional homogeneity of the sample aged for 240 h at 748 K.
This condition was selected to assess the influence of thermal
treatments on texture reduction in the alloy. From the EBSD
measurements, both inverse pole figures (IPFs) and pole figures (PFs)
were obtained. These representations use the unit m.r.d. (multiples of
uniform random distribution), which indicates how many times the
intensity of a given crystallographic orientation exceeds that of a
completely random and uniform distribution {[}25{]}, and guarantee that
the measurements were performed on texture-free samples.

\textbf{2.2 Neutron total scattering diffraction and Small-Angle Neutron
Scattering analysis}

Neutron total scattering and small-angle neutron scattering (SANS)
experiments were performed on CrCoNi samples for all conditions listed
in Table 1. All measurements were carried out at the Institut
Laue-Langevin (ILL), Grenoble, France.

Total scattering data were collected on the D4 two-axis diffractometer.
The incident beam size was 12 × 50 mm², and the wide-angle detector was
scanned over an angular range from 1.3$^\circ$ to 140$^\circ$, corresponding to a
\emph{Q}-range of 0.3--23.5 \AA$^{-1}$ at a wavelength of 0.499 \AA, thereby
providing access to high\emph{-Q} structural information. Two
experimental configurations were employed. The first involved
measurements of the empty sample holder, consisting of a cylindrical
container made of pure vanadium with an outer diameter of approximately
8 mm. The second comprised measurements of the CrCoNi samples sealed in
the same vanadium containers, ensuring experimental consistency and
enabling accurate background subtraction.

SANS data were acquired at the ILL using the D16 diffractometer. The instrument configuration limited the maximum scattering vector to $Q_{\mathrm{max}} = 0.06~\text{\AA}^{-1}$, thus probing structures larger than approximately $10~\mathrm{nm}$ in real space. After background subtraction, the intensities were normalized by the illuminated area, incident flux, attenuator settings, and wavelength ($\lambda = 4.5~\text{\AA}$, take-off angle $=85^\circ$). As usual in SANS, the scattering vector $Q$ is related to the real-space length scale through

\[
d=\frac{2\pi}{Q},
\]

where

\[
Q=\frac{4\pi}{\lambda}\sin\theta.
\]

\textbf{2.3 Synchrotron X-ray diffraction}

To compare the results obtained from neutron diffraction, three
additional samples were produced under controlled thermal-processing
conditions. Equiatomic CoCrNi alloys were synthesized by arc melting
high-purity elements (all \textgreater99.99\%) in an argon atmosphere.
All three samples were extracted from the same ingot, with a total mass
of approximately 50 g. The ingot was remelted five times to ensure
chemical homogeneity, then homogenized at 1473 K for 4 h, followed by
water quenching.

Subsequently, the material was hot-rolled at 1473 K to a thickness of 2
mm, cold-rolled to 50\% reduction, and recrystallized by annealing at
1273 K for 1 h, again followed by water quenching. Finally, the rolled
material was sectioned by electrical discharge machining (EDM) into
specimens with dimensions of approximately 6 mm in length, 4 mm in
width, and 1 mm in thickness for subsequent heat treatments.

After full recrystallization, two distinct aging treatments of 748 K for
100 h and 240 h were applied to promote different degrees of CSRO. In
total, three samples were analyzed: the annealed state (1273 K / 1 h),
and the two aged conditions (748 K / 100 h and 748 K / 240 h).

Synchrotron X-ray diffraction (SXRD) measurements were performed at the
DESY P07 beamline using a wavelength of 0.142346 \AA. A LaB₆ standard was
used to calibrate the sample-to-detector distance, detector tilt angle,
and beam center. A Perkin-Elmer detector with a resolution of 2048 ×
2048 pixels was employed for image acquisition, and the data were
integrated using the PyFAI software {[}26{]} with 1024 points,
converting the images into intensity versus 2$\theta$.

\textbf{2.4 Simulation Procedures}

The simulated neutron diffraction patterns were obtained from
independent Monte-Carlo (MC) modeling followed by molecular dynamics
(MD) computations. Following an established methodology {[}27{]},
equilibrium CSRO states were generated from independent atom-swap MC
simulations on 4000-atom FCC supercells. These simulations were carried
out using the Metropolis-Hasting algorithm {[}28{]}, with Cr, Co, and Ni
atoms initially distributed randomly on the FCC lattice. Each MC run was
initialized with a lattice constant previously equilibrated via NPT-MD
simulations using a Nosé--Hoover thermostat and barostat for 10,000
steps at the target temperature. During each MC simulation, 30 atom-swap
attempts per atom (60,000 total) were performed to approach chemical
equilibrium. This procedure was repeated for 300 K, 500 K, and 700 K to
obtain equilibrium CSRO states at each temperature, while the initial
configuration was used for the Random solid solution (RSS) state.

The final MC-equilibrated structures were then used as input for NPT-MD
simulations. These simulations were performed at a constant temperature
of 300 K, corresponding to room temperature of the diffraction
measurements, and zero hydrostatic pressure for 5,000 steps with a 2.5
fs timestep. Forty independent simulations, each containing 4,000 atoms,
were carried out for each CSRO state.

To mitigate finite-size effect in the atomistic simulations that could
lead to a high signal-to-noise ratio in neutron diffraction analysis,
supercells containing 32,000 atoms were constructed by combining 8
randomly selected snapshots from the 40 independent simulations of each
CSRO state. The resulting supercells were then relaxed at fixed volume
(i.e., energy minimized) before evaluating the neutron diffraction
patterns. The relaxed supercells were then analyzed along the z-axis to
evaluate the total projected intensity in this direction. Subsequently,
a two-dimensional Fourier Transform (FT-2D) was applied to obtain
diffraction patterns corresponding to the specific zone axes {[}110{]},
{[}112{]}, {[}111{]}, and {[}100{]}. The analyses were performed under
two distinct diffraction conditions: (\emph{i}) X-ray diffraction (XRD),
in which the atomic contrast was weighted by the atomic numbers of each
element, Cr (24), Co (27), and Ni (28); and (\emph{ii}) neutron
diffraction, using the neutron scattering lengths specific to each atom,
Cr (3.64 fm), Co (2.49 fm), and Ni (10.3 fm).

All simulations were performed using LAMMPS (Large-scale
Atomic/Molecular Massively Parallel Simulator) code {[}29{]}, employing
a Moment Tensor Potential {[}30{]} previously trained on high-fidelity
ab-initio data specifically curated to capture CSRO effects and their
impact on phase stability and related properties {[}23{]}.

\textbf{3. Results}

\textbf{3.1 Microstructural results}

Figure 1 presents the results obtained from SEM analysis. Figure 1a
shows secondary electron (SE) and backscattered electron (BSE) images at
different magnifications. These images indicate that, despite partial
sintering induced by the thermal treatments, individual powder particles
remain distinguishable and are not fully coalesced. Figure 1b displays
the EDS maps, confirming the compositional homogeneity of the sample.
Figures 1c present the inverse pole figure (IPF) map acquired via EBSD.
The corresponding PFs are shown in Figures 1d. A weak crystallographic
texture was observed, with maximum values below 1.5 m.r.d. These results
indicate that the homogenization and aging treatments effectively
minimized texture development in the material.

\includegraphics[width=\textwidth]{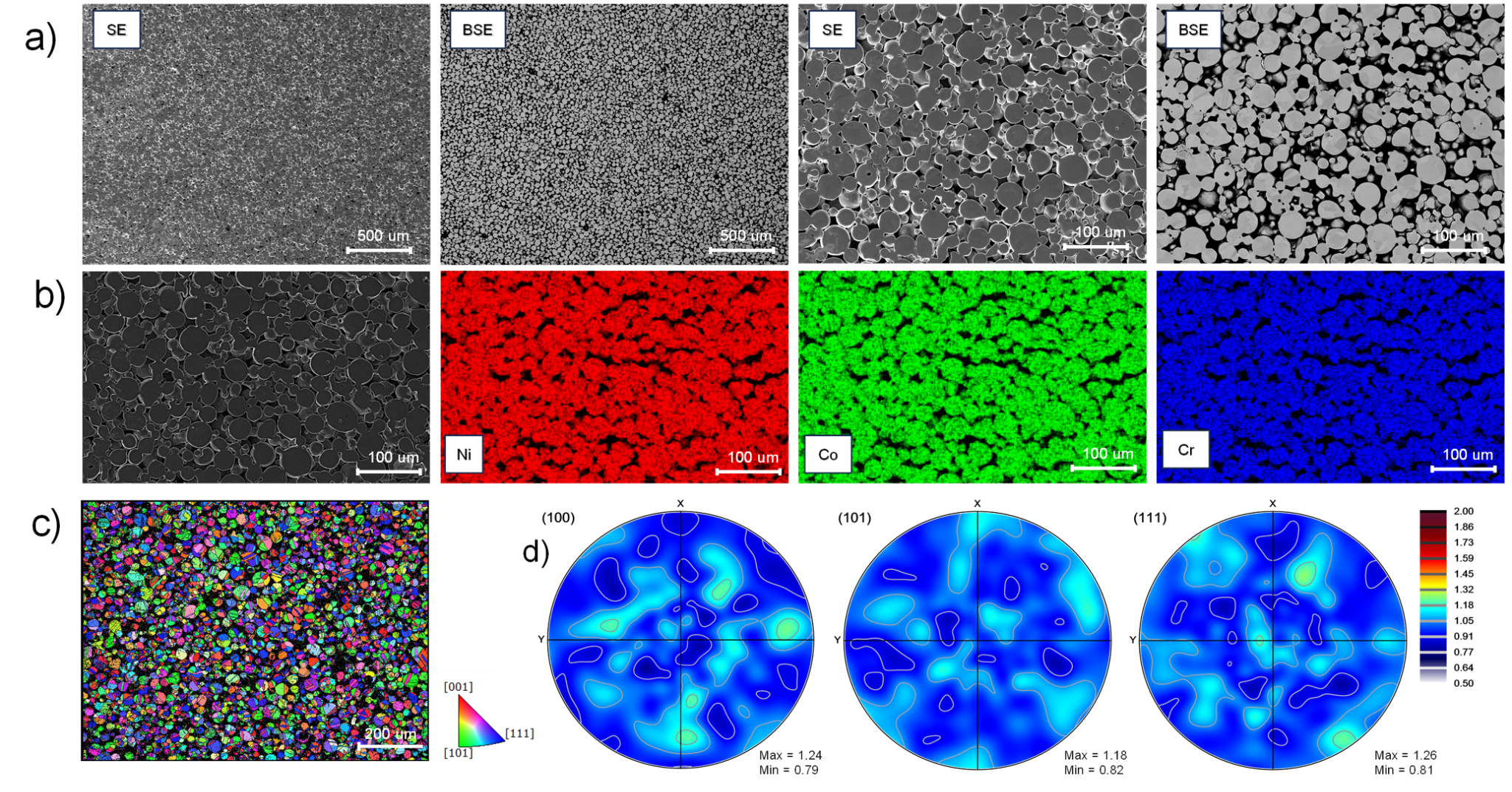}

Fig. 1 -- Scanning electron microscopy (SEM) analysis of the CoCrNi
alloy aged for 240 h at 748 K. (a) Secondary electron (SE) and
backscattered electron (BSE) images at different magnifications show
that, despite the partial sintering caused by the thermal treatments,
individual powder particles remain distinguishable and not fully
coalesced. (b) Energy-dispersive X-ray spectroscopy (EDS) maps confirm
the compositional homogeneity of the sample. (c) Inverse pole figure
(IPF) map acquired via electron backscatter diffraction (EBSD). (d)
Corresponding pole figures demonstrate that the applied processing
successfully reduced crystallographic texture.

\textbf{3.2 Diffraction analysis results}

Experimental neutron powder diffraction patterns are shown in Fig. 2.
All patterns were normalized by their maximum intensity to allow direct
comparison. The Bragg peaks are indistinguishable across all conditions,
confirming that the samples remain fully single-phase FCC with no
detectable changes in long-range crystallographic order. In contrast, a
distinct diffuse peak is observed in the low-\emph{Q} region
(highlighted by the red circle), centered near Q $\approx$ 1.85 \AA$^{-1}$.

The formation of CSRO is thermodynamically favored at lower
temperatures, where the reduced influence of entropy promotes local
ordering {[}31{]}. However, because CSRO develops through a diffusional
mechanism, its evolution at low temperatures requires longer aging times
to reach equilibrium {[}5{]}. Figure 2b shows an expanded version of the
region. For all samples, including the gas atomized sample, there
appears to be some form of diffuse feature present. With subsequent heat
treatment (either homogenization, or heat treatment), this diffuse
feature appears to change. Whilst there is an apparent change between
the three stages of sample preparation (gas-atomisation, homogenization,
and subsequent heat treatment), there is little significant change in
the observed between with the different post-homogenization heat
treatments.

\includegraphics[width=\textwidth]{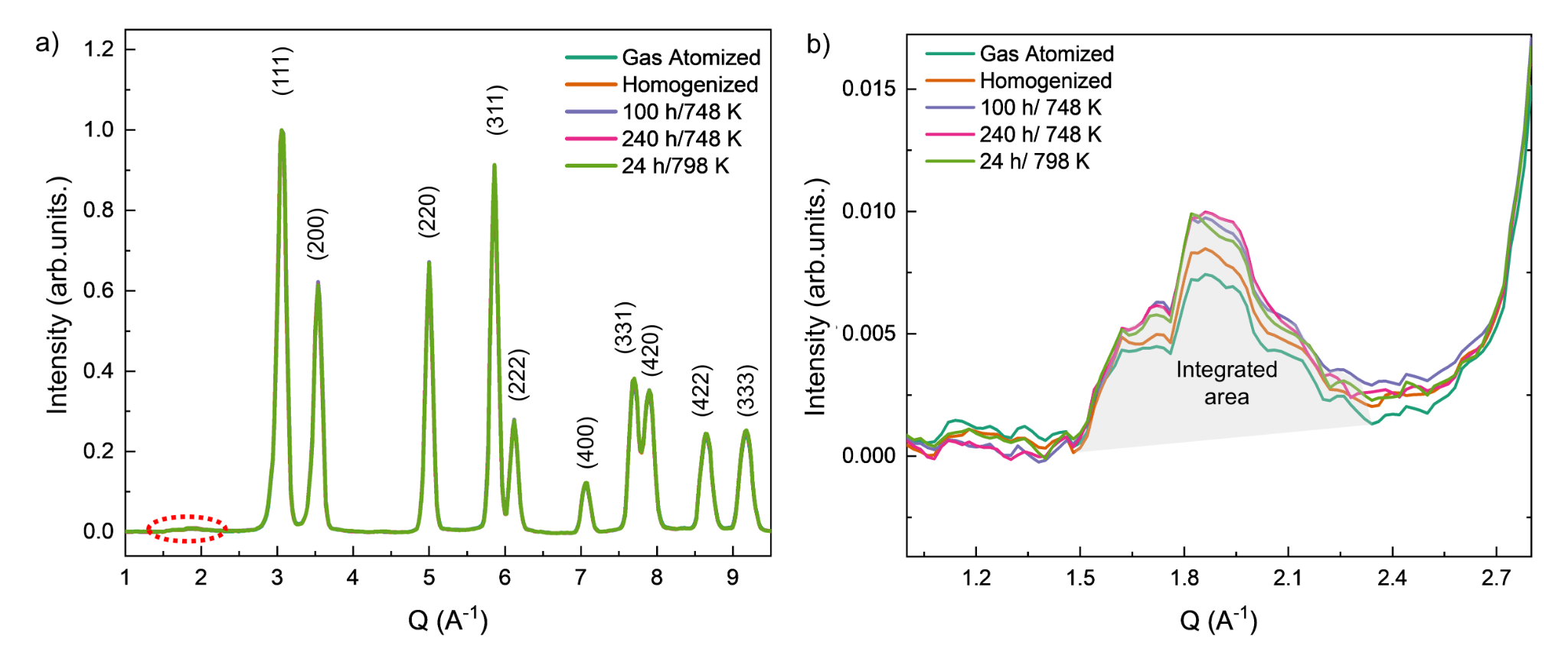}

Fig. 2 -- (a) Neutron powder diffraction patterns of CrCoNi under
different processing conditions. All patterns were normalized to their
respective maximum intensities. The Bragg peaks overlap perfectly for
all states (as-atomized, homogenized, and aged), indicating identical
average crystal structures and phase compositions. (b) The Q range
between 1 and 2.8 \AA$^{-1}$, where the most pronounced diffuse scattering
occurs. The diffuse peak reaches its maximum intensity at Q $\approx$ 1.85 \AA$^{-1}$,
corresponding to the position of forbidden (1 ½ 0) reflections
{[}22,32{]}. The gray-shaded region indicates the interval used to
measure the CSRO level in each sample.

To measure this effect, the area under the diffuse scattering peaks was
integrated over the same baseline (Fig. 2). The normalized areas for the
three aged samples are the highest, and are essentially identical. This
trend is consistent with calorimetric analyses {[}5{]}, which indicate
that CSRO saturation occurs after approximately 100 h of thermal
treatment at 748 K for CrCoNi. For the sample aged for a shorter time at
a higher temperature (24 h at 798 K), the observed behavior agrees with
earlier results {[}33{]}, which demonstrate through extensive electrical
resistivity measurements that CSRO saturation is achieved in
significantly shorter times when the temperature is increased to the
773--873 K range. Thus, it is reasonable to observe a comparable degree
of CSRO in the 24 h/798 K condition to that obtained in the 240 h/748 K
condition.

All results were normalized to the samples aged for 240 h at 748 K,
which were taken as the reference corresponding to the maximum CSRO
level achieved in this study. Remarkably, even the gas-atomized sample
displays approximately 70\% of the maximum diffuse intensity, indicating
that local chemical ordering is not suppressed by rapid solidification
{[}1,24{]}. This suggests that CSRO is a robust characteristic of the
CrCoNi system: it forms almost instantaneously upon quenching; it
persists across a wide range of processing routes; and it is further
enhanced by low-temperature aging {[}1{]}.

A similar trend was observed when analyzing the maximum intensity of the
diffuse peak, with the same aging conditions (100 h and 240 h at 748 K,
and 24 h at 798 K) exhibiting the highest values (Fig. 3). As with the
area under the peaks, all aged samples display CSRO levels that are very
close to each other, and within experimental uncertainty. This indicates
that, although neutron diffraction is sufficiently sensitive to detect
the presence of CSRO, it may not be able to resolve subtle differences
in magnitude among closely related aging conditions.

Considering that previous studies {[}5{]}, based on SXRD and TEM, have
demonstrated that these heat-treatment conditions are insufficient to
promote the formation of long-range ordered phases in equiatomic CoCrNi
alloys, the present observations can be attributed to CSRO effects.

To further evaluate whether the CSRO-related diffuse scattering at low Q
also affects the Bragg peaks associated with the FCC phase, both the
maximum intensity and the integrated area of each reflection were
analyzed (Supplementary Fig. 1a and 1b). These analyses revealed no
significant differences among the samples subjected to different
heat-treatment conditions, indicating that the FCC Bragg reflections
remain unchanged. This confirms that the observed effect is
intrinsically linked to the presence of CSRO, as the only detectable
variation occurs in the low-Q diffuse peak directly associated with the
development of this phenomenon.

A previous study {[}34{]} employing high-energy X-ray diffraction on
body-centered cubic (BCC) Fe--Ga alloys also reported the presence of
CSRO. In this work CSRO was identified through diffuse peaks at low
scattering angles, and the samples exhibiting the strongest diffuse
intensity displayed behavior analogous to that observed in the present
study.

\includegraphics[width=0.75\textwidth]{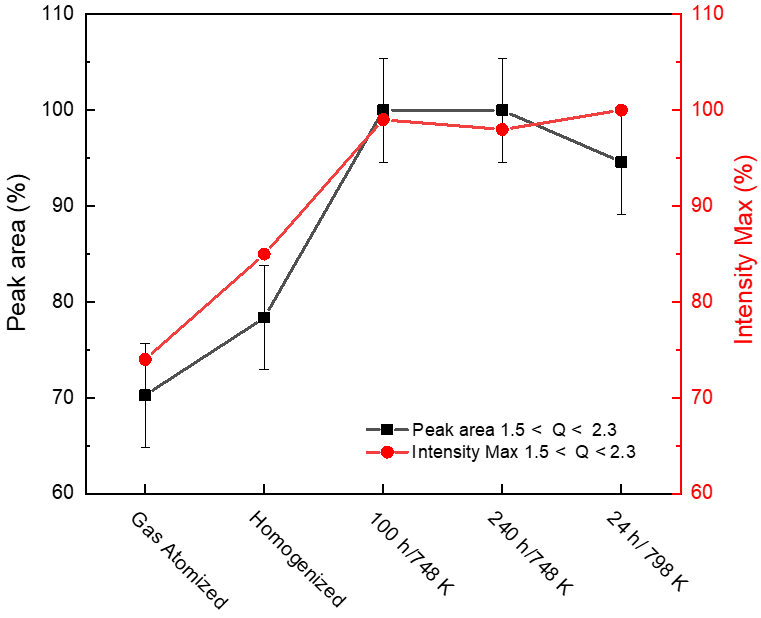}

Fig. 3 -- Measuring of CSRO in CrCoNi using the integrated area (black
symbols) and maximum intensity (red symbols) of the diffuse scattering
peak observed in neutron diffraction. The sample aged for 240 h at 748 K
was taken as the reference condition and assigned 100\% relative CSRO
content, with all other states normalized accordingly. Remarkably, the
as-atomized (gas-quenched) sample displays nearly 70\% of the maximum
CSRO, underscoring the speed with which local chemical correlations are
established. The uncertainty in the area calculation was estimated from
the quantified background noise, which was propagated to determine the
error associated with the integrated diffuse peak area.

To further investigate the presence and evolution of CSRO in the CrCoNi
alloy, we combined these neutron diffraction experiments with atomistic
simulations. Four systems were modeled: one representing an RSS with no
atomic ordering, and three additional configurations obtained by
equilibrating the system at 700 K, 500 K, and 300 K to generate
different levels of CSRO. All structures were fully relaxed, and their
corresponding diffraction patterns were subsequently computed (Fig. 4).
Consistent with earlier observations {[}29{]}, the degree of CSRO
increases systematically with decreasing equilibration temperature, with
the 300 K configuration exhibiting the highest level of ordering and the
700 K configuration the lowest.

To ensure a meaningful comparison with neutron diffraction measurements
acquired at room temperature, the MC-equilibrated structures associated
with different CSRO states were subsequently relaxed via NPT molecular
dynamics at 300 K and zero pressure. During this MD stage, atomic
identities were kept fixed, no atom swaps were allowed, so that the CSRO
levels obtained during MC equilibration were preserved. Although this
procedure does not explicitly model rapid cooling, it provides a
quenched configuration in which chemical correlations are frozen while
lattice parameters and atomic vibrations equilibrate to the experimental
measurement temperature. In this respect, the method mimics
experimentally quenched samples: the degree of CSRO reflects the
annealing temperature, whereas diffraction probes a room-temperature
structure.

Computed neutron diffraction patterns (intensity vs Q, Section 2.4) and
SXRD were compared directly with experimental measurements (Fig. 4). The
simulations reproduce the primary neutron-diffraction features,
including the diffuse peak between 1.5 \textless Q \textless 2.4 \AA$^{-1}$
(blue circle) and the FCC Bragg reflections (Fig. 4a). Conversely, SXRD
data show no indication of a diffuse peak at low Q. The weak features
appearing at 1.52 \AA$^{-1}$ and 1.76 \AA$^{-1}$, correspond to higher-order harmonics
of the FCC reflections, resulting from the absence of a monochromator
during the synchrotron experiment.

Figure 4b provides a magnified view of the 1.2 \textless Q \textless
2.8\,\AA$^{-1}$ range, where the diffuse scattering becomes more evident.
Except for the RSS configuration (orange dash dot-dot line), all
simulated structures exhibit a broad diffuse peak with maximum intensity
near Q $\approx$ 2.0\,\AA$^{-1}$, consistent with increased CSRO. The slight shift in
the position of this feature between simulated and experimental patterns
is attributed to differences in lattice parameters. As shown in
Supplementary Fig. 2, the lattice parameters from the simulations are
smaller than those measured experimentally, this is expected due to the
overbending issue typical of PBE exchange--correlation functionals (used
in the training of the machine learning potentials) {[}31{]}.
Consequently, the smaller lattice spacing in the simulations results in
diffraction features appearing at slightly higher \emph{Q} values.

\includegraphics[width=\textwidth]{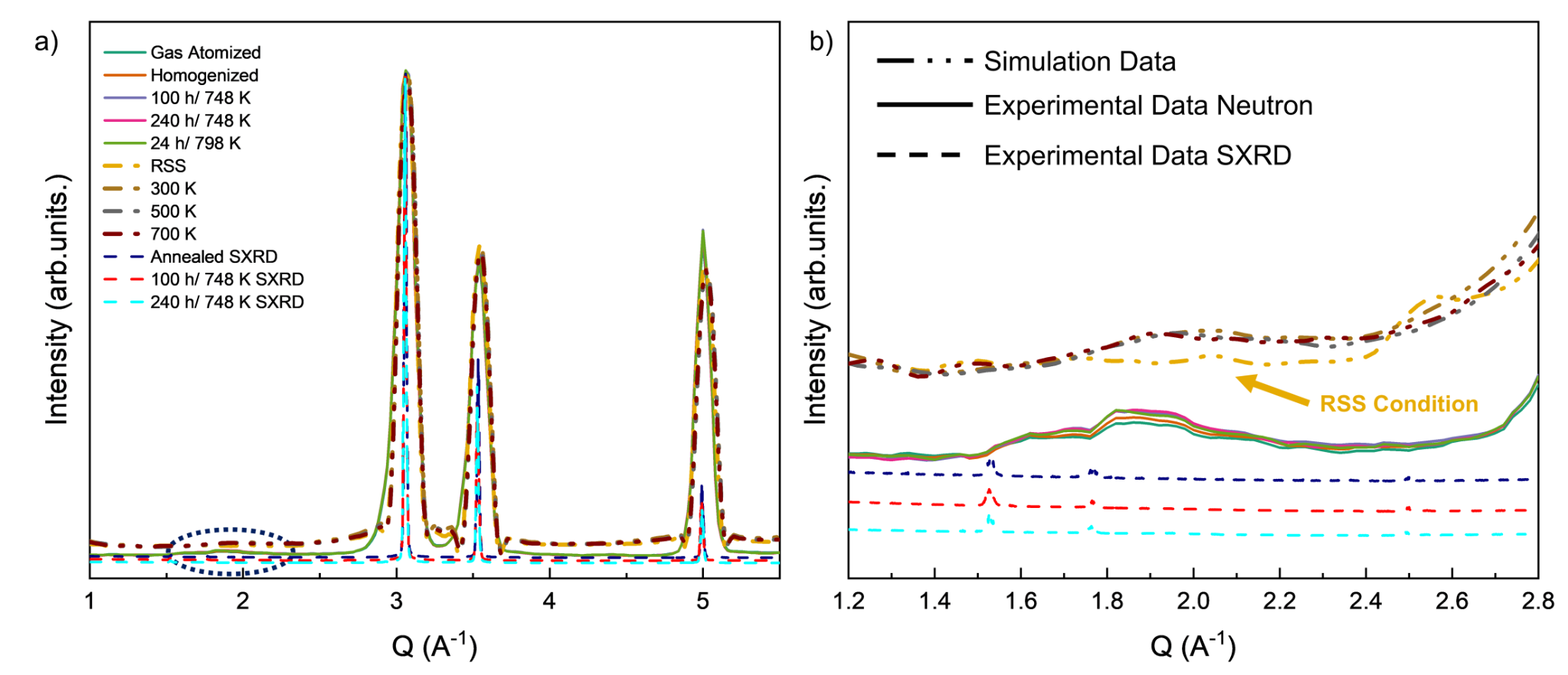}

\vspace{4mm}

Fig. 4 -- Comparison between experimental data and simulated neutron
diffraction patterns. (a) The overlaid curves reveal the presence of a
diffuse peak in the region 1.4 \textless Q \textless 2.4 \AA$^{-1}$ (blue
dotted circle), in addition to the Bragg peaks associated with the FCC
phase. (b) A magnified view of the highlighted region confirms that the
diffuse peak is reproduced in the simulated structures, demonstrating
strong agreement between experimental observations and first-principles
simulations. The SXRD data show no evidence of a diffuse peak,
confirming that this technique lacks sufficient scattering contrast
among Cr, Co, and Ni to resolve CSRO-related features. The simulated and
SXRD curves were vertically offset for improved clarity.

Figures 5a, 5c, 5e, and 5g correspond to the real-space projection of
the atomic structures along the z-axis, yielding simulated atomic
resolution images. These are analogous to high-resolution transmission
electron microscopy (HRTEM) images but with neutron scattering contrasts
instead, as the atomic contrast was weighted according to the neutron
scattering lengths of Cr (3.64 fm), Co (2.49 fm), and Ni (10.3 fm),
resulting in the contrast being dominated by Ni.

Although the visualization of arrays of atoms is difficult, Figures 5b,
5d, 5f, and 5h present the Fourier transforms of the atomic lattice
images for each condition. These can be understood as a proxy for the
single-crystal neutron diffraction pattern for each condition. Distinct
features arise depending on the degree of CSRO. For the RSS and 700~K
configurations, no significant diffuse scattering is detected between
the main Bragg reflections (Figs. 5b and 5d). In contrast, as CSRO
becomes more pronounced at lower temperatures, diffuse reflections
emerge at half-integer positions, such as (1 ½ 0) along the {[}100{]}
zone axis, highlighted by red circles in Figs. 4f and 4h. These features
are most prominent in the 300 K configuration, consistent with the
enhanced degree of chemical ordering.

\includegraphics[width=\textwidth]{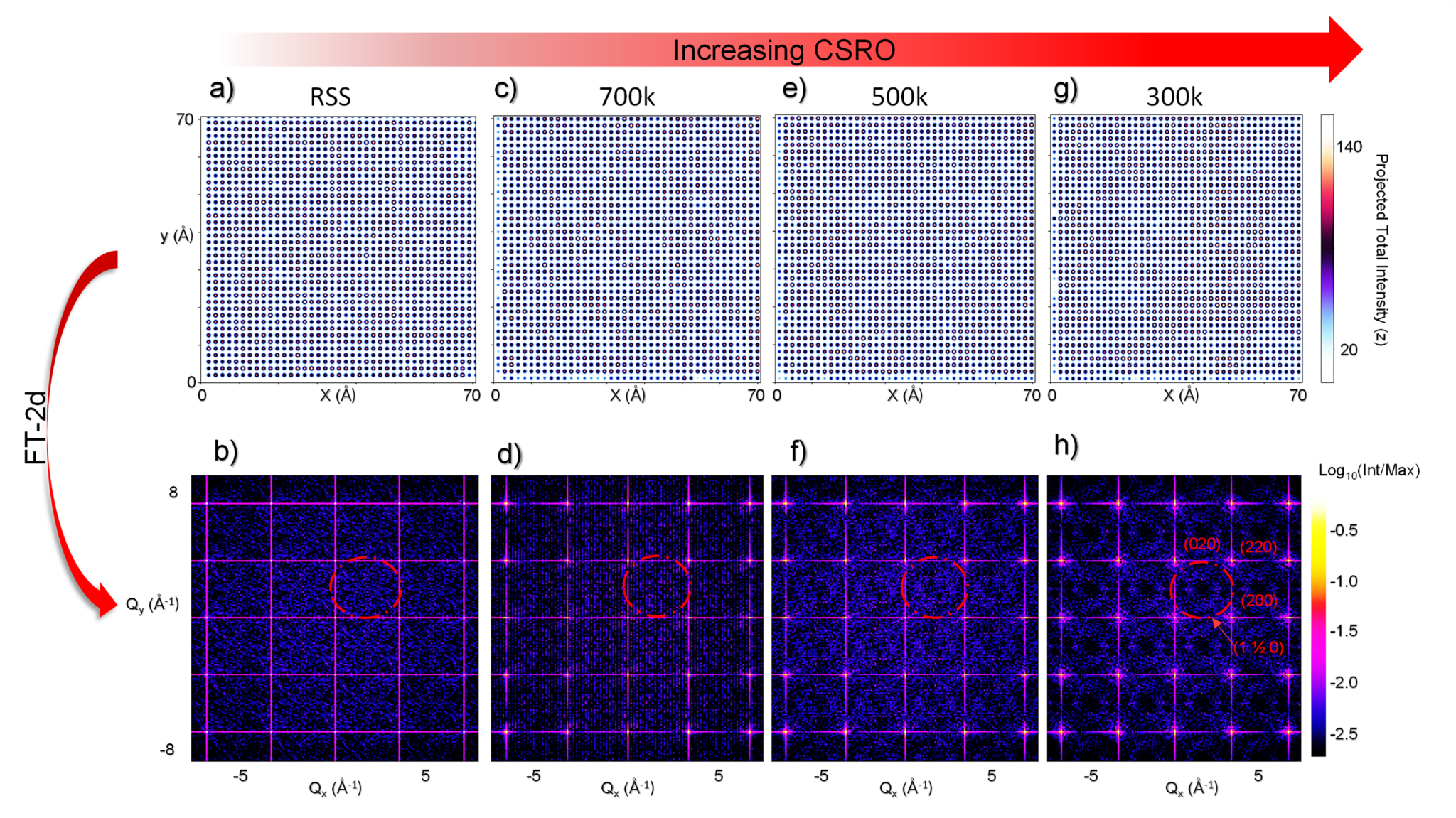}

Fig. 5 -- Simulations of atomic structures for different states. Panels
(a) and (b) correspond to a random solid solution (no CSRO); (c) and (d)
to a solid solution with CSRO annealed at 700 K; (e) and (f) to CSRO
annealed at 500 K; and (g) and (h) to CSRO annealed at 300 K. The top
images (a, c, e, g) show the real-space projection along the z-axis,
while the bottom (b, d, f, h) displays the corresponding two-dimensional
Fourier transforms (FT-2D), yielding the simulated neutron diffraction
patterns along the {[}100{]} zone axis. Red circles highlight the
diffuse spots associated with CSRO at the (1~½~0) position.

Based on the contrast observed in the FT-2D images, structural motifs
resembling D1a (Ni$_4$Mo prototype), Ni$_2$Cr-type ordering (Pt$_2$Mo prototype),
and D0$_2$$_2$ (Ni$_3$V prototype) structures could be identified. These features
have been previously reported in the literature through transmission
electron microscopy as diffuse spots along the {[}100{]} zone axis at
the (1 ½ 0) position, as well as in HRTEM images of compositions with
sufficient chemical contrast {[}32,35,36{]}. However, in the present
study, it was not possible to unambiguously determine which specific
structural motif is primarily responsible for the observed diffuse
spots. Notably, such features are absent along the other zone axes
examined (Supplementary Fig. 4), further supporting their direct
association with CSRO. In contrast, the presence of LRO would lead to
well-defined superlattice reflections appearing along multiple zone
axes, which is not observed in this case.

Notably, diffuse features are absent in the simulated X-ray diffraction
patterns, which were generated using the atomic scattering factors of Cr
(24), Co (27), and Ni (28) (Supplementary Fig. 5). This absence is
attributed to the low atomic number contrast among Cr, Co, and Ni for
X-ray scattering. These results are consistent with the experimental
SXRD data (Fig. 4), which likewise failed to detect the diffuse peak,
further confirming the limited sensitivity of X-rays (and thus
electrons) to resolve CSRO related features in this alloy system.
Moreover, along the {[}112{]} and {[}111{]} zone axis, no diffuse peaks
were detected, in agreement with the findings reported by Coury et al.
{[}16{]}. Their results indicate that the {[}112{]} orientation is not
the most reliable direction for detecting CSRO, in contrast to several
studies in the literature that have claimed CSRO signatures in
CrCoNi-based alloys using electron diffraction along this zone axis
{[}4,19,33,37{]}. Furthermore, the random supercell produces no
observable diffuse intensity, reinforcing the interpretation that the
half-order reflections are directly associated with CSRO.

The positions of the additional diffuse spots observed in the simulated
patterns (Fig. 4f and 4h) are located around \emph{Q} $\approx$ 1.85\,\AA$^{-1}$,
closely matching the diffuse peak identified in the experimental neutron
diffraction data shown in Fig. 2b. This strong agreement supports the
interpretation that the low-\emph{Q} diffuse scattering observed in the
experimental samples arises from CSRO. The simulated patterns confirm
that such features emerge in structures containing a high degree of
CSRO, further corroborating the experimental observations.

\textbf{3.2 Small-Angle Neutron Scattering analysis results}

The anomaly observed in the high-q SANS range, emerging near q $\approx$ 1.2 \AA$^{-1}$
(corresponding to a real-space length scale of \textasciitilde5.2 \AA), is
consistently present across all aged CrCoNi samples. This feature is
most likely an experimental artifact, as it coincides with the
transition to a second detector configuration (lower intensity,
increased statistical uncertainty). To mitigate its influence on the
quantitative SANS analysis, the dataset was cropped to q \textless
1.125 \AA$^{-1}$. Nevertheless, it is worth mentioning that this feature is
sometimes documented in Ni-based alloys with CSRO or severe lattice
distortions {[}14{]}. However, as shown in Supplementary Fig. 2, no
substantial variations in the lattice parameter are observed among the
samples, indicating that the applied heat treatments do not induce
measurable lattice distortions.

In the low-Q region, the scattering profile deviated from the classic
Porod law (I~∝~Q\textsuperscript{-4}), exhibiting a power-law exponent
between 1.9 and 2.5. According to the literature, an exponent of 2 is
often encountered in disordered structures, such as polymers' strands
coiled randomly; in the case of an ordered material, this exponent must
be linked to randomly oriented flat objects scattering centers
{[}38,39{]}. Current studies on steels have shown that an exponent near
two is associated with precipitates with the form factor of an oblate
spheroid or a thin disk {[}40{]}. The fine structure near Q $\approx$ 0.2
\AA\textsuperscript{-1} further suggests small compositional pockets or
fluctuations (Figure 6a). To reconcile these premises and observations,
we adopted a scattering model inspired in previous studies {[}41{]}
combining (1) a Porod term (exponent fixed as 3) to represent the
solid-solution, (2) a constant Laue background, (3) a lognormal
distribution of randomly oriented disks to account for two-dimensional CSRO domains,
and (4) an additional Lorentzian term to account for short-range
compositional heterogeneities (Equation 1). The Lorentzian is scaled by
the scattering length density contrast ($\Delta$$\rho$) to obtain a more robust fit.
Modeling and fit are performed with the aid of the jscatter library
{[}42{]}. The complete fitting with residuals is provided as
supplementary information Table S1.

I(Q) = Aq\textsuperscript{-3} + B + $\Delta$$\rho$\textsuperscript{2} ·
{[}vf\textsubscript{1} · D(Q) + vf\textsubscript{2} · S(Q){]} (1)

where:

• A = Porod prefactor;

• B = Laue background;

• D(Q) ≡ Polydisperse disks: lognormal distribution with $\sigma$ = 0.2,
adjusted R (radius) and D (thickness);

• S(Q) ≡ I\textsubscript{0}/
(1+$\xi$\textsuperscript{2}Q\textsuperscript{2}) (Lorentzian term with an
adjustable correlation length $\xi$)

• $\Delta$$\rho$ = SLD contrast = 4.312 × 10\textsuperscript{-4}
nm\textsuperscript{2},

• vf\textsubscript{1}, vf\textsubscript{2} = volume frac. scaling
factors

\includegraphics[width=\textwidth]{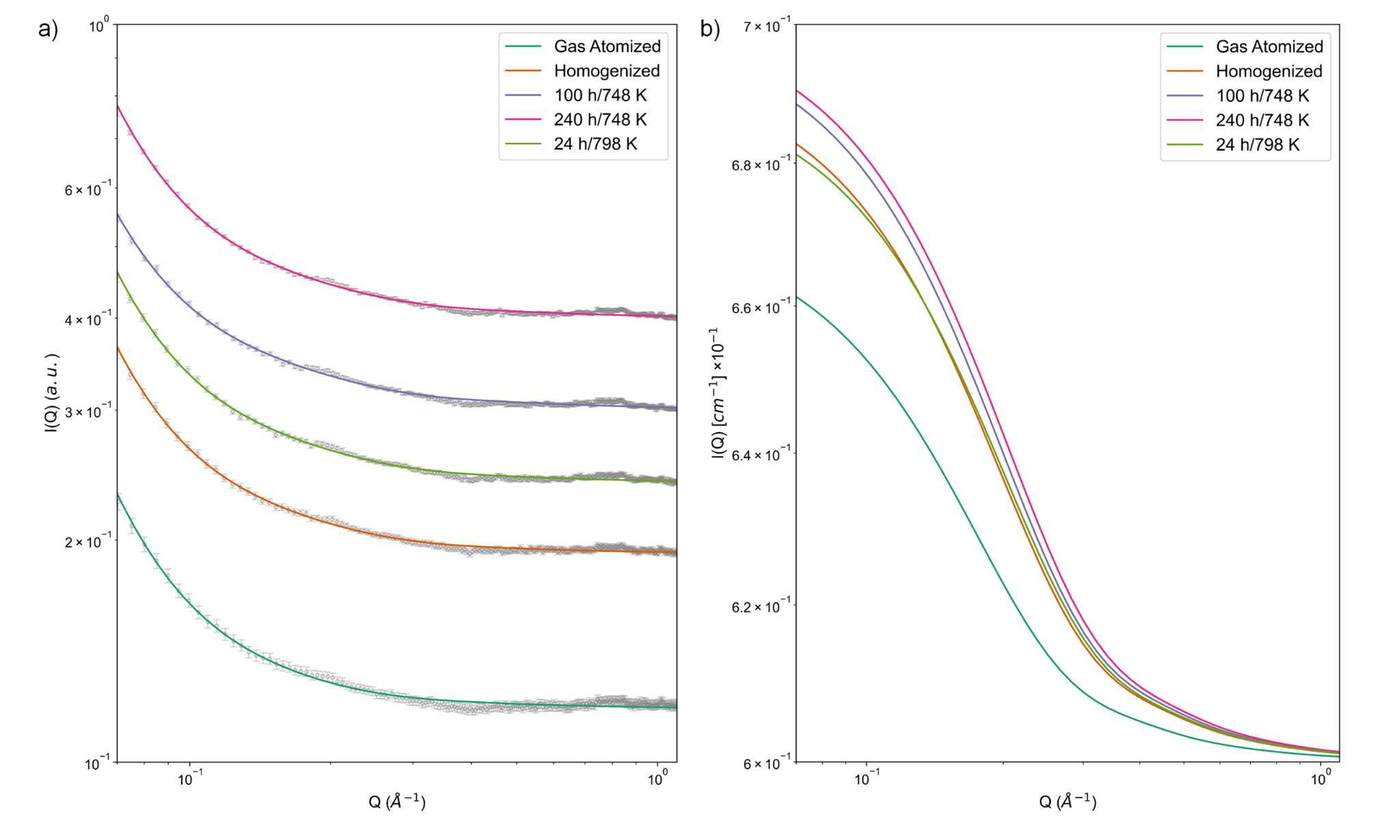}

Figure 6 - (a) SANS data from equiatomic CoCrNi samples subjected to
different processing and thermal treatments; the markers with error bars
(gray) indicate experimental data while the solid lines (colored)
represent the fitted models. (b) isolated contribution of SRO disks +
Lorentzian for scattering in each sample; the flat background was fixed
to 0.6. in the images, intensities are multiplied by
10\textsuperscript{6} for visualization purposes.

The neutron scattering length density (SLD) contrast was calculated
assuming four atoms per FCC unit cell with lattice parameter a = 3.564 ×
10\textsuperscript{-8}~cm. The ordered phase was considered pure Ni,
with a scattering length b = 1.03 × 10\textsuperscript{-12} cm. The
solid solution was assigned the average contrast of Ni, Co (b = 0.249 ×
10\textsuperscript{-12} cm), and Cr (b = 0.3635 ×
10\textsuperscript{-12} cm) {[}43{]}.

The fitting procedure yielded consistent geometric parameters for the
disk-like features, with radii of $\sim$ 11 \AA and thicknesses of $\sim$ 1 \AA ($\sigma$ =
0.2). The volumetric fraction of these disks, quantified as CSRO
domains, exhibits significant thermal activation. Relative to the gas
atomized powder (minimal vf\textsubscript{1} = 5.343 ×
10\textsuperscript{-6}), aging at 798 K for 24 h induced a 2.25-fold
increase in CSRO cluster density, while prolonged aging at 748 K (240 h)
further enhanced it to 2.63 times relative to the reference. These
clusters are at least one order of magnitude smaller than second-phase
precipitates usually analyzed via SANS {[}40{]}, but larger than the
radius of gyration Rg $\sim$ 3 \AA proposed by Hsiao et al. {[}14{]} under
similar conditions. This parameter describes the average size and
spatial distribution of atomic density within a cluster, providing a
quantitative measure of its nanoscale extent. The volume of CSRO based
on the scaling factor of disks (vf\textsubscript{1}) is as follows:
aging at 748 K (240 h \textgreater 100 h) \textgreater aging at 798
K (24 h) \textgreater Homogenized \textgreater Gas atomized. The
volume fraction of disk-shaped features (vf1) and other best-fit
parameters, along with their associated uncertainties, are provided in
Supplementary Table S1. It is worth noting that, even in the absence of
precipitate formation in the samples, the SANS profiles differ
significantly from one another, indicating variations in the atomic
arrangement.

The Lorentzian term improves the fitting stability, but the correlation
length ($\xi$) is saturated at the minimum (bound) of 1.0 \AA, which is
comparable to the atomic radius of Ni ($\sim$ 1.24 \AA), thus too small to
associate with CSRO. At the same time, the vf\textsubscript{2} scaling
factor follows the same trend as vf\textsubscript{1}. Relative to the
homogenized reference sample (minimal vf\textsubscript{2} = 3.56 \%),
vf\textsubscript{2} increases by a factor of 1.33 in the sample aged at
798 K and by 2.03 in the sample aged at 748 K (100 h). Such a behavior
is highlighted in Fig. 6b. It is worth mentioning that despite
vf\textsubscript{2} being 10\textsuperscript{4} larger than
vf\textsubscript{1}, the disks dominate scattering due to their form
factor scaling as ($\pi$R\textsuperscript{2}D)\textsuperscript{2} (i.e.,
R\textsuperscript{4}D\textsuperscript{2}), with R and D being the radius
and the thickness of disks, while the Lorentzian term scales weakly with
$\xi$\textsuperscript{3}.

\begin{enumerate}
\def\labelenumi{\arabic{enumi}.}
\setcounter{enumi}{3}
\item
  \textbf{Discussion}
\end{enumerate}

\textbf{4.1 Origin of Diffuse Spots Along the {[}100{]} Zone Axis at the
(1 ½ 0) Position}

CSRO in metallic alloys typically manifests as broad and diffuse
intensity maxima in reciprocal space {[}32{]}. These maxima occur at
points where the second derivative of the free energy reaches a minimum,
locations that, for symmetry reasons, satisfy the Lifshitz criterion,
which requires that two or more symmetry elements intersect at the same
position in reciprocal space {[}44{]}. The main distinction between CSRO
and LRO lies in the nature of the additional reflections. In LRO,
well-defined superlattice peaks appear at multiple reciprocal-space
positions and across different zone axes, reflecting the periodicity of
the ordered phase. In contrast, in the present work, CSRO is
characterized by the presence of diffuse reflections, typically located
at positions of the type (1 ½ 0) along the {[}100{]} zone axis in FCC
structures. These features arise from local atomic ordering associated
with the \{420\} family of planes {[}32,36{]}.

The occurrence of diffuse spots at (1 ½ 0), observed along the {[}100{]}
zone axis, has been widely reported in alloys with an FCC structure and
is invariably attributed to the presence of CSRO. Van Tendeloo and
Amelinckx {[}32{]} investigated Ni$_4$Mo, Au$_4$Cr, Au$_4$V, and Au$_4$Fe alloys
using HRTEM after heat treatments carefully designed to suppress the
formation of long-range ordered phases. In these systems, the chemical
contrast between constituent elements is sufficient to enable the
detection of CSRO via transmission electron microscopy, which is not the
case for the CrCoNi alloy. In all instances, diffuse reflections at (1 ½
0) were identified prior to second phase precipitation, although with
varying intensities, more pronounced in Ni$_4$Mo and weaker in Au$_4$Fe. The
authors interpreted these reflections as structural projections
associated with the D1\textsubscript{a} and D0\textsubscript{22} phases.

Similar results were later reported by Hata et al. {[}36{]} in the
Ni--19.5\% Mo alloy, also analyzed by HRTEM. Using Fourier
transformation with a doughnut-shaped aperture, the authors confirmed
the presence of diffuse spots at (1 ½ 0) along the {[}100{]} axis. After
background noise reduction in the HRTEM images, patterns consistent with
the structural projections D1\textsubscript{a}, D0\textsubscript{22},
Pt$_2$Mo, and N\textsubscript{2}M\textsubscript{2} were identified, the
latter being embedded in the FCC matrix. The observation of
N\textsubscript{2}M\textsubscript{2} patterns was consistent with the
maximum intensities detected at (1 ½ 0), while mixed configurations of
D1\textsubscript{a}, D0\textsubscript{22}, and Pt$_2$Mo were also reported
in other studies {[}35,45,46{]}, indicating that at least part of these
projections contributes to the origin of the diffuse reflections. In the
same work, the experimental results were corroborated by Monte Carlo
simulations, which reproduced similar diffuse patterns, further
reinforcing the connection between the (1 ½ 0) reflections and the
presence of CSRO.

In the present work, and in comparison with the studies mentioned above,
the FT-2D patterns derived from the simulated structures also revealed
diffuse reflections at the (1 ½ 0) position, consistent with the
projection of the D1a structure observed in simulated FFT patterns
{[}32{]}. It is worth noting that the simulated neutron-based images
were generated from monocrystalline models, which may contribute to the
reduced visibility of other possible structural motifs, such as D0$_2$$_2$ and
Pt$_2$Mo, as reported in previous studies. Based on these simulated
structures, the manifestation of CSRO in equiatomic CrCoNi can be
associated with reflections arising from projections related to D1a-,
Pt$_2$Mo-, and D0$_2$$_2$-type motifs. However, it is not possible to
unambiguously identify the dominant configuration, and further
experimental investigations on single-crystal samples are required to
reach a definitive conclusion. Furthermore, only minimal contrast is
expected in electron and X-ray diffraction patterns along the {[}100{]}
zone axis for CrCoNi, due to the very similar structure factors of Cr,
Co, and Ni. This explains why such diffuse features have not been
observed using these techniques (as shown in Supplementary Figs. 5 and
6).

\textbf{4.2 SANS discussion}

The SANS models proposed in this work align well with recent atom probe
tomography (APT) data published by Li et al. {[}47{]}. Their study
identified excess Ni-Ni bonds within clusters ranging from 20 to 155
atoms (equivalent to diameters of 0.7--1.5 nm), consistent with our disk
sizes (2.2 nm diameter). Furthermore, while the authors mention in the
main text that most (number density) CSRO domains are spherical, their
supplementary data reveal occasional 2D Ni-Ni clusters (disks or rods)
that typically contain up to 300 atoms. Given that scattered intensity
scales with the square of the volume, even a small population of such
bidimensional clusters would dominate the SANS signal, as proposed
herein. It is important to note that the scattering length of Ni is
significantly different from Co and Cr, which are similar to one
another, thus, the neutron diffraction data will display enhanced
contrast when Ni-based features are present. Thus, there might be other
types and shapes of CSRO-related domains in this alloy but these are the
most prominent regarding Ni.

\begin{enumerate}
\def\labelenumi{\arabic{enumi}.}
\setcounter{enumi}{4}
\item
  \textbf{Conclusions}
\end{enumerate}

Neutron diffraction results revealed variations in diffuse intensity
scattering near Q $\approx$ 1.85 \AA$^{-1}$. Simulated structures generated by MD/MC
methods reproduced diffuse scattering at low angles, analogous to the
experimental observations, suggesting CSRO as the origin of these
signals.

Based on these results, the extent of CSRO was measure by integrating
the area under the diffuse scattering peak. Samples aged at 748 K for
100 h and 240 h, as well as the sample aged at the slightly higher
temperature of 798 K for 24 h, exhibited a significantly higher degree
of CSRO compared to the as-atomized and homogenized conditions.
Remarkably, even the rapidly solidified powder displayed approximately
70\% of the CSRO level observed in the sample aged for 240 h, indicating
that this ordering phenomenon is quick enough to stabilize the phase
even during rapid solidification.

Analysis of the simulated structures through \emph{FT}-2D revealed the
presence of diffuse reflections along the {[}100{]} zone axis at the (1
½ 0) positions, features commonly reported in other FCC alloys
exhibiting CSRO. These reflections can be associated with structural
motifs such as D1a, Pt$_2$Mo, and D0$_2$$_2$, as described in the literature.

The SANS results indicate that the contribution of Ni-rich domains
(disks or rods) to neutron scattering cannot be neglected. The disk form
factor implies that CSRO manifests as localized, plate-like
compositional fluctuations rather than isotropic clusters, aligning with
observations from atom-probe tomography.

Neutron diffraction is a powerful technique for detecting CSRO in
CrCoNi, overcoming the limitations of transmission electron diffraction,
where the low contrast among Cr, Co, and Ni hinders direct observation
of the phenomenon. As such, the neutron-based approach establishes a
robust and sensitive pathway for identifying and measuring CSRO in this
system.

\textbf{Acknowledgment:} V.P.B. was supported by Fundação de Amparo à
Pesquisa do Estado de São Paulo - Brasil (FAPESP) under grants
2024/02515-2 To CAPES - Coordenação de Aperfeiçoamento de Pessoal de
Nível Superior for their financial support in this work through a
scholarship, process no. 88887.952714/2024-00. This study was financed
in part by the Coordenação de Aperfeiçoamento de Pessoal de Nível
Superior - Brasil (CAPES) - Finance Code 001. G.C.S. was supported by
Fundação de Amparo à Pesquisa do Estado de São Paulo - Brasil (FAPESP)
under grant 2023/07403-5. A.F.A. acknowledges the financial support from
FAPESP, grant 2024/11008-7 and the National Council for Scientific and
Technological Development {[}CNPQ, grants 157806/2025-1 and
153727/2024-1{]}. E.M.M. acknowledges Conselho Nacional de
Desenvolvimento Científico e Tecnológico (CNPq) processes 444505/2024-5
and 403997/2025-9. F.G.C. acknowledges the Fundação de Amparo à Pesquisa
do Estado de São Paulo (FAPESP), process number 2022/02770-7, and the
Conselho Nacional de Desenvolvimento Científico e Tecnológico (CNPq),
processes 403832/2023-3 and 444393/2024-2, for the financial support.
The research was sponsored by the Army Research Office and was
accomplished under Grant Number W911NF-23--1--0310. The views and
conclusions contained in this document are those of the authors and
should not be interpreted as representing the official policies, either
expressed or implied, of the Army Research Office or the U.S.
Government. The U.S. Government is authorized to reproduce and
distribute reprints for Government purposes notwithstanding any
copyright notation herein. The authors thank the Laboratory of
Structural Characterization (LCE/ DEMa/UFSCar) for the general
facilities. We acknowledge DESY (Hamburg, Germany), a member of the
Helmholtz Association HGF, for the provision of experimental facilities.
Parts of this research were carried out at PETRA III and we would like
to thank Drs. Norbert Schell, Emad Maawad, and Martin Etter for
assistance in using the P07 beamline. The authors would like to
sincerely acknowledge the Institut Laue-Langevin (ILL), Grenoble,
France, for providing the neutron beamtime and experimental facilities.
We are especially grateful to Dr. Gabriel Julio Cuello for his valuable
support and assistance during the experiments performed on the D4
instrument, as well as to Dr. Viviana Cristiglio for her essential help
during the experiments carried out on the D16 instrument. Their
expertise and guidance were fundamental to the successful completion of
this work.

\textbf{Author contributions: CRediT} V.P.B., F.G.C., and E.M.M.
conceptualized this research and designed the entire workflow. V.P.B.
and G.C.S. conducted all the production and processing of the alloys
used in the study. V.P.B., C.B.S., C.S., and L.O. conducted the neutron
total scattering diffraction and small-angle neutron scattering
analyses. F.G.C. conducted synchrotron radiation diffraction. R.F. and
Y.C. conducted the MD-MC analyses. V.P.B., G.C.S., and F.G.C. processed
data obtained from neutron total scattering diffraction. C.S. processed
data obtained from small-angle neutron scattering analyses. V.P.B.,
F.G.C., and E.M.M. drafted the first manuscript. A.F.A., L.O. and D.M.
revised the manuscript.

\textbf{Declaration of Competing Interests:} The authors declare that
they have no known competing financial interests or personal
relationships that could have appeared to influence the work reported in
this paper.

\textbf{Data and materials availability:} The data supporting the
findings in this study are available within the paper. Any further
information or clarification is available from the corresponding author
upon reasonable request.

\textbf{References}

{[}1{]} Y. Han, H. Chen, Y. Sun, J. Liu, S. Wei, B. Xie, Z. Zhang, Y.
Zhu, M. Li, J. Yang, W. Chen, P. Cao, Y. Yang, Ubiquitous short-range
order in multi-principal element alloys, Nat. Commun. 15 (2024) 6486.
https://doi.org/10.1038/s41467-024-49606-1.

{[}2{]} W.H. Blades, B.W.Y. Redemann, N. Smith, D. Sur, M.S. Barbieri,
Y. Xie, S. Lech, E. Anber, M.L. Taheri, C. Wolverton, T.M. McQueen, J.R.
Scully, K. Sieradzki, Tuning chemical short-range order for stainless
behavior at reduced chromium concentrations in multi-principal element
alloys, Acta Mater. 277 (2024) 120209.
https://doi.org/10.1016/j.actamat.2024.120209.

{[}3{]} X. Chen, Q. Wang, Z. Cheng, M. Zhu, H. Zhou, P. Jiang, L. Zhou,
Q. Xue, F. Yuan, J. Zhu, X. Wu, E. Ma, Direct observation of chemical
short-range order in a medium-entropy alloy, Nature. 592 (2021)
712--716. https://doi.org/10.1038/s41586-021-03428-z.

{[}4{]} L. Zhou, Q. Wang, J. Wang, X. Chen, P. Jiang, H. Zhou, F. Yuan,
X. Wu, Z. Cheng, E. Ma, Atomic-scale evidence of chemical short-range
order in CrCoNi medium-entropy alloy, Acta Mater. 224 (2022) 117490.
https://doi.org/10.1016/j.actamat.2021.117490.

{[}5{]} V.P. Bacurau, P.A.F.P. Moreira, G. Bertoli, A.F. Andreoli, E.
Mazzer, F.F. de Assis, P. Gargarella, G. Koga, G.C. Stumpf, S.J.A.
Figueroa, M. Widom, M. Kaufman, A. Fantin, Y. Cao, R. Freitas, D.
Miracle, F.G. Coury, Comprehensive analysis of ordering in CoCrNi and
CrNi2 alloys, Nat. Commun. 15 (2024) 7815.
https://doi.org/10.1038/s41467-024-52018-w.

{[}6{]} E.P. George, D. Raabe, R.O. Ritchie, High-entropy alloys, Nat.
Rev. Mater. 4 (2019) 515--534.
https://doi.org/10.1038/s41578-019-0121-4.

{[}7{]} Y. Zhang, T.T. Zuo, Z. Tang, M.C. Gao, K.A. Dahmen, P.K. Liaw,
Z.P. Lu, Microstructures and properties of high-entropy alloys, Prog.
Mater. Sci. 61 (2014) 1--93.
https://doi.org/10.1016/j.pmatsci.2013.10.001.

{[}8{]} B. Gludovatz, A. Hohenwarter, K.V.S. Thurston, H. Bei, Z. Wu,
E.P. George, R.O. Ritchie, Exceptional damage-tolerance of a
medium-entropy alloy CrCoNi at cryogenic temperatures, Nat. Commun. 7
(2016) 10602. https://doi.org/10.1038/ncomms10602.

{[}9{]} B. Gwalani, T. Alam, C. Miller, T. Rojhirunsakool, Y.S. Kim,
S.S. Kim, M.J. Kaufman, Y. Ren, R. Banerjee, Experimental investigation
of the ordering pathway in a Ni-33 at.\%Cr alloy, Acta Mater. 115 (2016)
372--384. https://doi.org/10.1016/j.actamat.2016.06.014.

{[}10{]} S.M. Dubiel, J. Cieslak, Short-range order in iron-rich Fe-Cr
alloys as revealed by Mössbauer spectroscopy, Phys. Rev. B. 83 (2011)
180202. https://doi.org/10.1103/PhysRevB.83.180202.

{[}11{]} B. Stephan, D. Jacob, F. Delabrouille, L. Legras, A kinetic
study of order-disorder transition in ni--cr based alloys, Miner. Met.
Mater. Ser. (2019) 233--249.
https://doi.org/10.1007/978-3-030-04639-2\_15.

{[}12{]} Y. Liu, H. Lou, F. Zhang, Z. Zeng, Q. Zeng, Short-range order
in binary and multiple principal element alloys: A review, Matter
Radiat. Extrem. 10 (2025). https://doi.org/10.1063/5.0275123.

{[}13{]} F.X. Zhang, S. Zhao, K. Jin, H. Xue, G. Velisa, H. Bei, R.
Huang, J.Y.P. Ko, D.C. Pagan, J.C. Neuefeind, W.J. Weber, Y. Zhang,
Local Structure and Short-Range Order in a NiCoCr Solid Solution Alloy,
Phys. Rev. Lett. 118 (2017) 205501.
https://doi.org/10.1103/PhysRevLett.118.205501.

{[}14{]} H.-W. Hsiao, R. Feng, H. Ni, K. An, J.D. Poplawsky, P.K. Liaw,
J.-M. Zuo, Data-driven electron-diffraction approach reveals local
short-range ordering in CrCoNi with ordering effects, Nat. Commun. 13
(2022) 6651. https://doi.org/10.1038/s41467-022-34335-0.

{[}15{]} H. Joress, B. Ravel, E. Anber, J. Hollenbach, D. Sur, J.
Hattrick-Simpers, M.L. Taheri, B. DeCost, Why is EXAFS for complex
concentrated alloys so hard? Challenges and opportunities for measuring
ordering with X-ray absorption spectroscopy, Matter. 6 (2023)
3763--3781. https://doi.org/10.1016/j.matt.2023.09.010.

{[}16{]} F.G. Coury, C. Miller, R. Field, M. Kaufman, On the origin of
diffuse intensities in fcc electron diffraction patterns, Nature. 622
(2023) 742--747. https://doi.org/10.1038/s41586-023-06530-6.

{[}17{]} F. Walsh, M. Zhang, R.O. Ritchie, M. Asta, A.M. Minor, Multiple
origins of extra electron diffractions in fcc metals, Sci. Adv. 10
(2024). https://doi.org/10.1126/sciadv.adn9673.

{[}18{]} L.R. Owen, H.Y. Playford, H.J. Stone, M.G. Tucker, A new
approach to the analysis of short-range order in alloys using total
scattering, Acta Mater. 115 (2016) 155--166.
https://doi.org/10.1016/j.actamat.2016.05.031.

{[}19{]} R. Zhang, S. Zhao, J. Ding, Y. Chong, T. Jia, C. Ophus, M.
Asta, R.O. Ritchie, A.M. Minor, Short-range order and its impact on the
CrCoNi medium-entropy alloy, Nature. 581 (2020) 283--287.
https://doi.org/10.1038/s41586-020-2275-z.

{[}20{]} A.F. Andreoli, A. Fantin, S. Kasatikov, V.P. Bacurau, M. Widom,
P. Gargarella, E.M. Mazzer, T.G. Woodcock, K. Nielsch, F.G. Coury, The
impact of chemical short-range order on the thermophysical properties of
medium- and high-entropy alloys, Mater. Des. 238 (2024) 112724.
https://doi.org/10.1016/j.matdes.2024.112724.

{[}21{]} F. Walsh, M. Asta, R.O. Ritchie, Magnetically driven
short-range order can explain anomalous measurements in CrCoNi, Proc.
Natl. Acad. Sci. U. S. A. 118 (2021) 1--6.
https://doi.org/10.1073/pnas.2020540118.

{[}22{]} W. Schweika, H.G. Haubold, Neutron-scattering and Monte Carlo
study of short-range order and atomic interaction in Ni0.89Cr0.11, Phys.
Rev. B. 37 (1988) 9240--9248. https://doi.org/10.1103/PhysRevB.37.9240.

{[}23{]} Y. Cao, K. Sheriff, R. Freitas, Capturing short-range order in
high-entropy alloys with machine learning potentials, (2024) 40--42.
http://arxiv.org/abs/2401.06622.

{[}24{]} M. Islam, K. Sheriff, Y. Cao, R. Freitas, Nonequilibrium
chemical short-range order in metallic alloys, Nat. Commun. 16 (2025)
8926. https://doi.org/10.1038/s41467-025-64733-z.

{[}25{]} A.I. Saville, S.C. Vogel, A. Creuziger, J.T. Benzing, A.L.
Pilchak, P. Nandwana, J. Klemm-Toole, K.D. Clarke, S.L. Semiatin, A.J.
Clarke, Texture evolution as a function of scan strategy and build
height in electron beam melted Ti-6Al-4V, Addit. Manuf. 46 (2021)
102118. https://doi.org/10.1016/j.addma.2021.102118.

{[}26{]} J. Kieffer, D. Karkoulis, PyFAI, a versatile library for
azimuthal regrouping, J. Phys. Conf. Ser. 425 (2013) 202012.
https://doi.org/10.1088/1742-6596/425/20/202012.

{[}27{]} K. Sheriff, Y. Cao, T. Smidt, R. Freitas, Quantifying chemical
short-range order in metallic alloys, Proc. Natl. Acad. Sci. 121 (2024).
https://doi.org/10.1073/pnas.2322962121.

{[}28{]} N. Metropolis, A.W. Rosenbluth, M.N. Rosenbluth, A.H. Teller,
E. Teller, Equation of State Calculations by Fast Computing Machines, J.
Chem. Phys. 21 (1953) 1087--1092. https://doi.org/10.1063/1.1699114.

{[}29{]} S. Plimpton, Fast Parallel Algorithms for Short-Range Molecular
Dynamics, J. Comput. Phys. 117 (1995) 1--19.
https://doi.org/10.1006/jcph.1995.1039.

{[}30{]} A. V. Shapeev, Moment Tensor Potentials: A Class of
Systematically Improvable Interatomic Potentials, Multiscale Model.
Simul. 14 (2016) 1153--1173. https://doi.org/10.1137/15M1054183.

{[}31{]} F. Coury, G. Stumpf, Y. Cao, V. Bacurau, D. Miracle, W. Wolf,
E. Zanotto, R. Freitas, On the nature of chemical short-range order
evolution, (2025). https://www.researchsquare.com/article/rs-7339299/v1.

{[}32{]} G. Van Tendeloo, S. Amelinckx, D. de Fontaine, On the nature of
the ``short-range order'' in 1 1/2 0 alloys, Acta Crystallogr. Sect. B
Struct. Sci. 41 (1985) 281--292.
https://doi.org/10.1107/S0108768185002166.

{[}33{]} L. Li, Z. Chen, S. Kuroiwa, M. Ito, K. Yuge, K. Kishida, H.
Tanimoto, Y. Yu, H. Inui, E.P. George, Evolution of short-range order
and its effects on the plastic deformation behavior of single crystals
of the equiatomic Cr-Co-Ni medium-entropy alloy, Acta Mater. 243 (2023)
118537. https://doi.org/10.1016/j.actamat.2022.118537.

{[}34{]} S. Guruswamy, T. V. Jayaraman, R.P. Corson, G. Garside, S.
Thuanboon, Short range ordering and magnetostriction in Fe-Ga and other
Fe alloy single crystals, J. Appl. Phys. 104 (2008).
https://doi.org/10.1063/1.3040154.

{[}35{]} K.H. Lee, K. Hiraga, D. Shindo, M. Hirabayashi, High resolution
electron microscopic study of the ordering processes in Ni4Mo alloy,
Acta Metall. 36 (1988) 641--649.
https://doi.org/10.1016/0001-6160(88)90098-3.

{[}36{]} S. Hata, S. Matsumura, N. Kuwano, K. Oki, D. Shindo, Short
range order in Ni4Mo and its high resolution electron microscope images,
Acta Mater. 46 (1998) 4955--4961.
https://doi.org/10.1016/S1359-6454(98)00180-3.

{[}37{]} M. Zhang, Q. Yu, C. Frey, F. Walsh, M.I. Payne, P. Kumar, D.
Liu, T.M. Pollock, M.D. Asta, R.O. Ritchie, A.M. Minor, Determination of
peak ordering in the CrCoNi medium-entropy alloy via nanoindentation,
Acta Mater. 241 (2022) 118380.
https://doi.org/10.1016/j.actamat.2022.118380.

{[}38{]} C.J. Gommes, S. Jaksch, H. Frielinghaus, Small-angle scattering
for beginners, J. Appl. Crystallogr. 54 (2021) 1832--1843.
https://doi.org/10.1107/S1600576721010293.

{[}39{]} J. Eyssautier, P. Levitz, D. Espinat, J. Jestin, J. Gummel, I.
Grillo, L. Barré, Insight into Asphaltene Nanoaggregate Structure
Inferred by Small Angle Neutron and X-ray Scattering, J. Phys. Chem. B.
115 (2011) 6827--6837. https://doi.org/10.1021/jp111468d.

{[}40{]} Y.Q. Wang, S.J. Clark, V. Janik, R.K. Heenan, D.A. Venero, K.
Yan, D.G. McCartney, S. Sridhar, P.D. Lee, Investigating
nano-precipitation in a V-containing HSLA steel using small angle
neutron scattering, Acta Mater. 145 (2018) 84--96.
https://doi.org/10.1016/j.actamat.2017.11.032.

{[}41{]} G. Spartacus, J. Malaplate, F. De Geuser, I. Mouton, D. Sornin,
M. Perez, R. Guillou, B. Arnal, E. Rouesne, A. Deschamps, Chemical and
structural evolution of nano-oxides from mechanical alloying to
consolidated ferritic oxide dispersion strengthened steel, Acta Mater.
233 (2022) 117992. https://doi.org/10.1016/j.actamat.2022.117992.

{[}42{]} R. Biehl, Jscatter, a program for evaluation and analysis of
experimental data, PLoS One. 14 (2019) e0218789.
https://doi.org/10.1371/journal.pone.0218789.

{[}43{]} V.F. Sears, Neutron scattering lengths and cross sections,
Neutron News. 3 (1992) 26--37.
https://doi.org/10.1080/10448639208218770.

{[}44{]} D. De Fontaine, Configurational Thermodynamics of Solid
Solutions, in: 1979: pp. 73--274.
https://doi.org/10.1016/S0081-1947(08)60360-4.

{[}45{]} P. De Meulenaere, G. Van Tendeloo, J. Van Landuyt, D. Van Dyck,
On the interpretation of HREM images of partially ordered alloys,
Ultramicroscopy. 60 (1995) 265--282.
https://doi.org/10.1016/0304-3991(95)00065-9.

{[}46{]} A. Verma, N. Wanderka, J.B. Singh, M. Sundararaman, J. Banhart,
On the evolution of long-range order from short-range order in a Ni
2(Cr0.5Mo0.5) alloy, J. Alloys Compd. 586 (2014) 561--566.
https://doi.org/10.1016/j.jallcom.2013.10.086.

{[}47{]} Y. Li, T. Colnaghi, Y. Gong, H. Zhang, Y. Yu, Y. Wei, B. Gan,
M. Song, A. Marek, M. Rampp, S. Zhang, Z. Pei, M. Wuttig, S. Ghosh, F.
Körmann, J. Neugebauer, Z. Wang, B. Gault, Machine Learning‐Enabled
Tomographic Imaging of Chemical Short‐Range Atomic Ordering, Adv. Mater.
36 (2024). https://doi.org/10.1002/adma.202407564.

\end{document}